\documentclass[preprints,article,accept,moreauthors,pdftex]{Definitions/mdpi} 

\usepackage{bm}
\usepackage{bbm}
\usepackage{listings}
\usepackage{subfig}
\usepackage{amsmath}
\usepackage{amssymb}
\usepackage{booktabs}
\usepackage{hyperref}
\usepackage{amsthm}
\hypersetup{colorlinks,citecolor=blue}

\firstpage{1} 
\pubvolume{xx}
\issuenum{1}
\articlenumber{1}
\pubyear{2020}
\copyrightyear{2020}
\history{}

\Title{The Pricing of Vanilla Options with Cash Dividends as a Classic Vanilla Basket Option Problem}
\newcommand*\diff{\mathop{}\!\kern0pt\mathrm{d}}

\Author{Jherek Healy}
\AuthorNames{Jherek Healy}
\address{}
\corres{Correspondence: jherekhealy@protonmail.com}

\abstract{
	In the standard Black-Scholes-Merton framework, dividends are represented as a continuous dividend yield and the pricing of Vanilla options on a stock is achieved through the well-known Black-Scholes formula. In reality however, stocks pay a discrete fixed cash dividend at each dividend ex-date. This leads to the so-called piecewise lognormal model, where the asset jumps from a fixed known amount at each dividend date. There is however no exact closed-form formula for the pricing of Vanilla options under this model. Approximations must be used. While there exists many approximations taylored to this specific problem in the litterature, this paper explores the use of existing well-known basket option formulas for the pricing of European options on a single asset with cash dividends in the piecewise lognormal model.
}

\begin{document}
	
\section{Introduction}

\section{The piecewise-lognormal model}
We assume that the stock price $S$ follows a lognormal process between dividend dates, and jumps at each dividend ex-date $t_i$ from the cash dividend amount $\alpha_i$. The corresponding stochastic differential equation reads

\begin{equation}\label{eqn:spot_model}
	\begin{cases}
		\diff S = r_R(t)S(t) \diff t + \sigma(t) S(t) \diff W \textmd{, } t_{i} \leq t < t_{i+1}\,,\\
		S(t_{i}) = S(t_{i}^{-}) - \alpha_i\,,
	\end{cases}
\end{equation}
where $(t_i)_i$ denotes the full ordered set of dividend dates, $r_R$ is the stock lending rate (also known as repo rate), and $W$ is a Brownian motion.

The price of a European call option of strike $K$, maturity $T$  on $S$ reads
\begin{align}
	V_C(T) &= B(0,T_p)\mathbb{E}_{Q}\left[ \max(S(T) - K , 0)\right]\,,
\end{align}
where $B(0,T_p)$ is the discount factor to the payment date (typically two business days after the maturity date), calculated using the risk free rate, or the collateral rate. Let $\tilde{V}_C = \frac{V_C}{B(0,T_p)}$ be the undiscounted option price.
Integrating Equation \ref{eqn:spot_model} leads then to
\begin{align}
\tilde{V}_C(T)&=\mathbb{E}_\mathbb{Q}\left[\max\left( 	S(0) e^{\int_{0}^T r_R(s) - \frac{\sigma^2(s)}{2} \diff s + \int_0^{T} \sigma(s) \diff W(s) } - \sum_{i=1}^n \alpha_i e^{ \int_{t_i}^T r_R(s) - \frac{\sigma^2(s)}{2} \diff s + \int_{t_i}^{T} \sigma(s) \diff W(s) } - K,0 \right)\right]\,,
\end{align}
where $n$ is the number of dividends whose ex-date falls between the valuation date and the maturity date.


\section{Recasting the problem in terms of a basket option}
Let 
\begin{align}
	\tilde{W}(s) &= W(t) - \int_{0}^t \sigma(s) ds\,,\\
	M(t) &= e^{\int_{0}^{t} \sigma(s) \diff W(s) - \frac{1}{2} \int_{0}^t \sigma^2(s) \diff s }\,.
\end{align}
We have
\begin{equation*}
	\tilde{V}_C(T)=C(0,T)\mathbb{E}_\mathbb{Q}\left[M(T) \max\left( 	S(0)   - \sum_{i=1}^n \frac{\alpha_i e^{ \int_{0}^{t_i}  \frac{\sigma^2(s)}{2} \diff s - \int_{0}^{t_i} \sigma(s) \diff W(s) }}{C(0,t_i)} - \frac{K}{M(T)C(0,T)} ,0 \right)\right]\,,
\end{equation*}
with $C(0,T) = e^{\int_{0}^T r_R(s) \diff s }$.
The Girsanov theorem \cite[Theorem 5.2.3]{Shreve04} tells us that $\tilde{W}$ is a Brownian motion under the probability measure $\tilde{Q}$ defined by $M$. The Radon-Nikodym theorem \cite[Lemma 5.2.2]{Shreve04} leads to 
\begin{equation}\label{eqn:cash_basket}
	\tilde{V}_C(T)=C(0,T)\mathbb{E}_{\tilde{Q}}\left[\max\left( 	S(0)   - \sum_{i=1}^n \frac{\alpha_i e^{ -\int_{0}^{t_i}  \frac{\sigma^2(s)}{2} \diff s + \int_{0}^{t_i} \sigma(s) \diff \tilde{W}(s) }}{C(0,t_i)}  - \frac{K e^{-\int_{0}^{T}\frac{\sigma^2(s)}{2} \diff s + \int_0^{T} \diff \tilde{W}(s) }}{C(0,T)} ,0 \right)\right]\,.
\end{equation}

The expectation in Equation \ref{eqn:cash_basket} corresponds to the undiscounted price under the standard Black-Scholes model of a European put option of strike $S(0)$  and maturity $T$ on a basket composed of the lognormal assets $(S_i)$ with drift $\mu_i$ and volatility $\sigma_i$ defined by
\begin{align}
	S_i(0) &= \alpha_i\,, \\
	\mu_i &= -\frac{1}{T}\int_0^{t_i} r_R(s) \diff s\,,\\
	\sigma_i^2  &= \frac{1}{T}\int_0^{t_i} \sigma^2(s) \diff s\,,
\end{align}
for $1 \leq i \leq n $. 

If $t_n=T$, we let $S_n = \alpha_n + K$ instead of $S_n = \alpha_n$ and the basket in composed of $n$ assets. Otherwise, we define $S_{n+1} = K, \mu_{n+1} = -\frac{1}{T}\int_0^{T} r_R(s) \diff s, \sigma_{n+1}^2 = -\frac{1}{T}\int_0^{T} \sigma^2(s) \diff s$ and the basket is composed of $n+1$ assets.

Thus, we have shown that any classic basket option approximation may be used to price European options on a stock paying discrete cash dividends, under the piecewise lognormal model.

\section{Remarks}
\subsection{Dividend policy}
In absence of a dividend policy, the piecewise lognormal model allows for negative stock prices, if $\alpha_i > S(t_j^-)$.
\citet{haug2003back} explore two natural choices : the liquidator policy where $S(t_i) = \max(0, S(t_i^-)-\alpha_i)$ and the survivor policy where $S(t_i) = \max(S(t_i^-), S(t_i^-) -\alpha_i)$.

The liquidator dividend policy does not impact the call option price, because we have
\begin{align*}
 \max\left( \max\left(	S(0)   - \sum_{i=1}^n \frac{\alpha_i e^{ \int_{0}^{t_i}  \frac{\sigma^2(s)}{2} \diff s - \int_{0}^{t_i} \sigma(s) \diff W(s) }}{C(0,t_i)},0\right) - \frac{K}{M(T)C(0,T)} ,0 \right) =\\\  \max\left( 	S(0)   - \sum_{i=1}^n \frac{\alpha_i e^{ \int_{0}^{t_i}  \frac{\sigma^2(s)}{2} \diff s - \int_{0}^{t_i} \sigma(s) \diff W(s) }}{C(0,t_i)} - \frac{K}{M(T)C(0,T)} ,0 \right)
	\end{align*}
for $K \geq 0$.

It will however impact the put option price $V_P$, and put options must then be priced using the put-call parity formula $V_C - V_P = B(0,T_p)(F(0,T) - K)$ with forward equal to the undiscounted call option price of strike zero $F(0,T)=\tilde{V}_C(T, K=0)$.

\subsection{Affine dividends}
In addition to cash dividends, we may include proportional dividends where $S$ jumps as $S(t_j^+)= S(t_j^-) - \beta_j S(t_{j}^{-})$. The formulae will not change, except for the definition of $C(0,T)$ which then must include the proportional dividends as follows
\begin{equation}
	C(0,T) = e^{\int_0^T r_R(s) \diff s} \prod_{j: 0 < t_j \leq T} (1-\beta_j)\,.
\end{equation}

\section{Numerical tests}
We assess the accuracy of the basket approach against the modern and accurate cash dividend approximation of \citet{sahel2011matching}, the second-order approximation of \citet{zhang2011fast} and the third-order formula of \citet{lefloch2015more} respectively named "GS", "Zhang-2" and "LL-3". As basket option approximation, we evaluate the one of \citet{ju2002pricing} named "BB-Ju" the more standard Curran based approximation of \citet{deelstra2010moment} named "BB", and the related lower bound approximation named "BB-LB" in the figures and tables.

\citet{deelstra2010moment} present several variations. While those do not significantly change the results, using their notation, we choose $\delta_i = \delta_3 = e^{r(T-j)}$  and $f_s(\Lambda) = f_3(\Lambda) = F\mathbb{G}_F$ as it led to the best results in \cite[Figures 1 and 2]{deelstra2010moment}.

\subsection{Single dividend case}
We consider a European call option of maturity one year on a stock with spot price $S=100$, and a single dividend $\alpha_1 = 7$, varying the ex-date, for three different strikes. To stay in line with the setup of \citet{haug2003back}, we let the interest rate be $r=6\%$ and the volatility $\sigma=30\%$. We look at the error in terms of Black-Scholes implied volatility. This is a good scale to compare
out-of-the-money with at-the-money or in-the-money options as well as to compare options of short maturity with options of long maturity. This is also a very natural measure in the context of the volatility surface construction.

\begin{figure}[h]
	\begin{center}  
		
			\includegraphics[width=\textwidth]{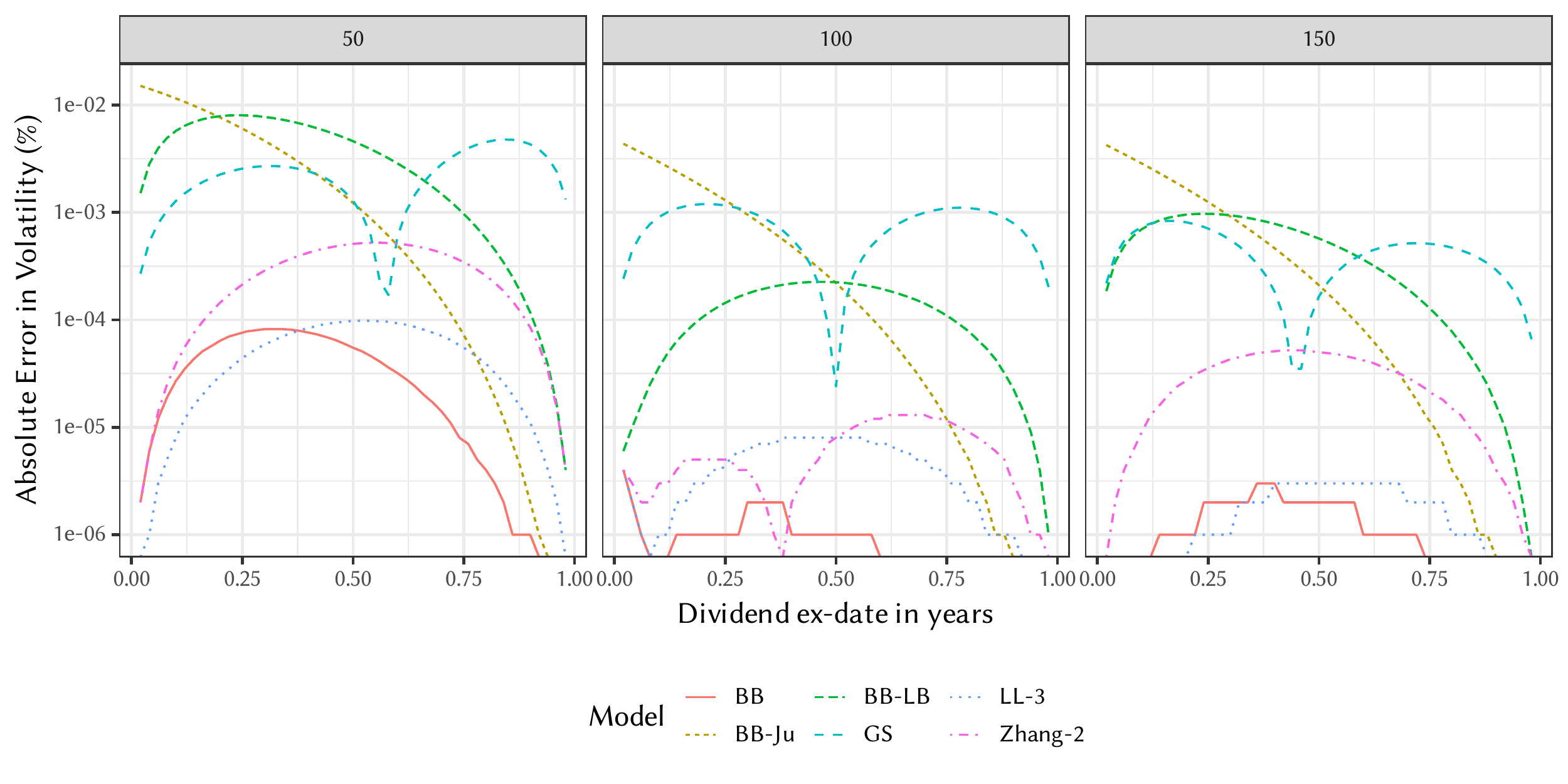}		
		\caption{Error of the various approximation on the case of the single dividend $\alpha_1=7$, varying the dividend ex-date for strike $K=50, K=100$ and $K=150$.\label{fig:single_div_td}}
	\end{center}
\end{figure}

The Ju approximation is very accurate when the dividend ex-date is close to maturity, but exhibits a large error when the ex-date is close the valuation date (Figure \ref{fig:single_div_td}). 
The error of the Deelstra approximation is much more contained, and is largest when the dividend falls at $t=0.4$. Overall, the error of the basket approach with the Deelstra approximation is found to be the most accurate, more accurate than the third-order method of Le Floc'h. The formula from Sahel and Gocsei is around two orders of magnitudes less accurate on this single dividend example. The lower bound basket approximation is competitive with the formula of Sahel and Gocsei in terms of accuracy.

\subsection{Many dividends case}
We reproduce the example of \citet{sahel2011matching}, of a call option of maturity 10 years, on a stock of price $S=100$, paying dividends of amount $\alpha_i=2$ on a semi-annual schedule, with the first dividend starting one day from the valuation date at $t_1 = 1/365$. The interest rate is taken to be $r=3\%$ and the volatility $\sigma=25\%$.

\begin{figure}[h]
	\begin{center}  
		
		\includegraphics[width=\textwidth]{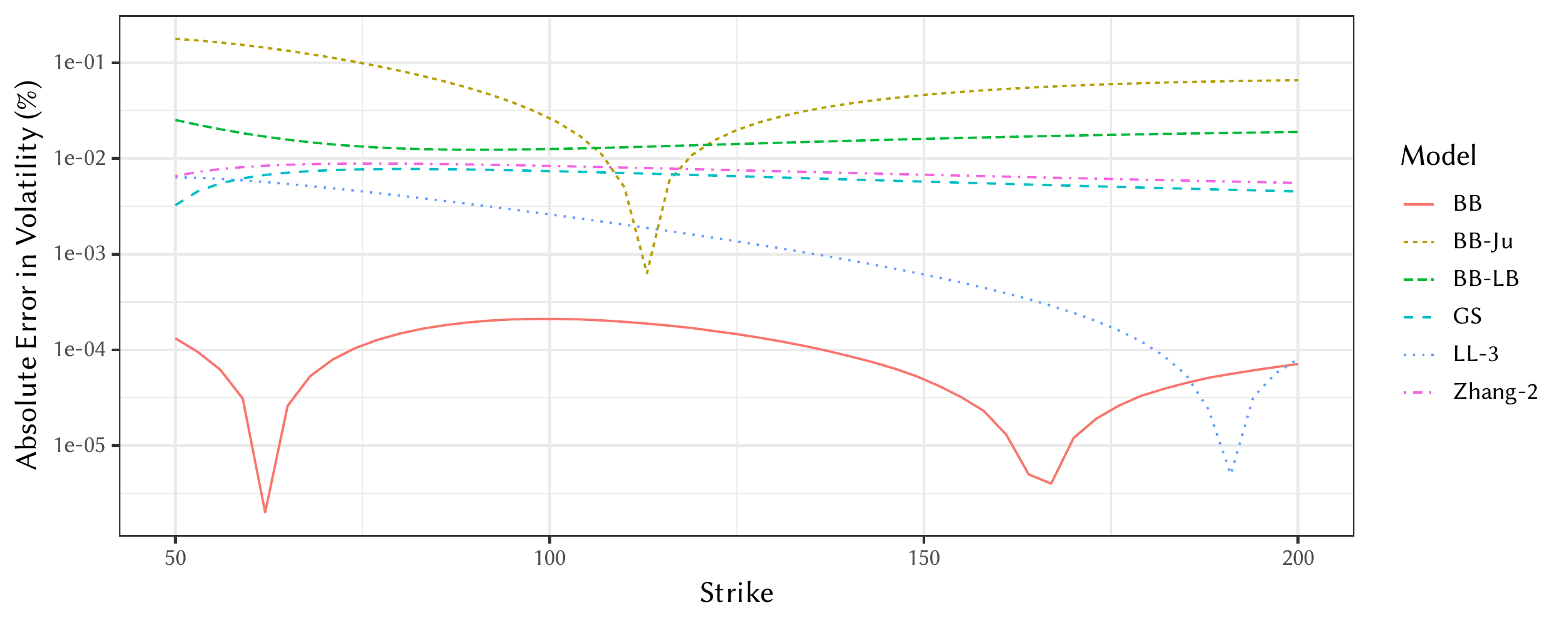}		
		\caption{Error of the various approximation in the case of semi-annual dividends $\alpha_i=2$, for an option of maturity 10 years, varying the strike price.\label{fig:gocsei_10y}}
	\end{center}
\end{figure}

Figure \ref{fig:gocsei_10y} shows that the basket approach, with a  maximum absolute error of $0.0002$ volatility point, is the most accurate. 
With a maximum error of 0.0078 volatility point, GS is slightly more accurate than Zhang-2, the basket approach with the Ju formula leads to the largest error and the lower bound approximation leads to a maximum error of 0.0252 volatility points at the lowest strike. 

In Table \ref{tbl:vellekoop}, we reproduce the example of a 7-year option, with varying yearly dividend amounts, first presented in \cite{vellekoop2006efficient} and in \cite{etore2012stochastic}. Our reference, named "FDM" is the price obtained by the TR-BDF2 finite difference method using 10,000 space steps and 3,650 time-steps and is accurate to the fifth decimal. We also give the values of the Richardson extrapolated binomial tree from \citet{vellekoop2006efficient}, named "VNRE", and of the third order approximation from \citet{etore2012stochastic}, named "EG-3".
\begin{table}[h]
	\centering{
		\caption{Price of a call option of maturity 7 years, on an asset with spot price $S=100$ paying yearly dividends of respective amounts 6, 6.5, 7, 7.5 8, 8, 8, starting at $t_1=0.9$, with $r=6\%$ and $\sigma=25\%$. \label{tbl:vellekoop}}
	\begin{tabular}{l c c c c c c c c}\toprule
		Strike & FDM & VNRE & EG-3 & LL-3 & GS & BB-Ju & BB & BB-LB\\\midrule
		70 &  27.21395 & 27.21 & 27.20498 & 27.21768 & 27.26457 & 27.18748 & 27.21394 & 27.20683\\
		100 & 19.48229 & 19.48 & 19.47843 & 19.48478 & 19.51519 & 19.47156 & 19.48245 & 19.47511\\
		130 & 4.13026 & 14.13 & 14.12933 & 14.13139 & 14.15233 & 14.13643 & 14.13045 & 14.12143\\ \midrule
		MRE & N/A & N/A & 3E-4 & 1E-4 & 2E-3 & 8E-4 & 1E-5 & 6E-4\\
		\bottomrule
	\end{tabular}}
\end{table}
On this example, the maximum relative error in price (MRE) is largest for GS ($2\cdot10^{-3}$), and smallest for BB (around $1\cdot10^{-5}$, at the limit of our reference price accuracy obtained  by the TR-BDF2 finite difference method).
\subsection{Extreme settings}
\begin{figure}[h]
	\begin{center}  
		\includegraphics[width=\textwidth]{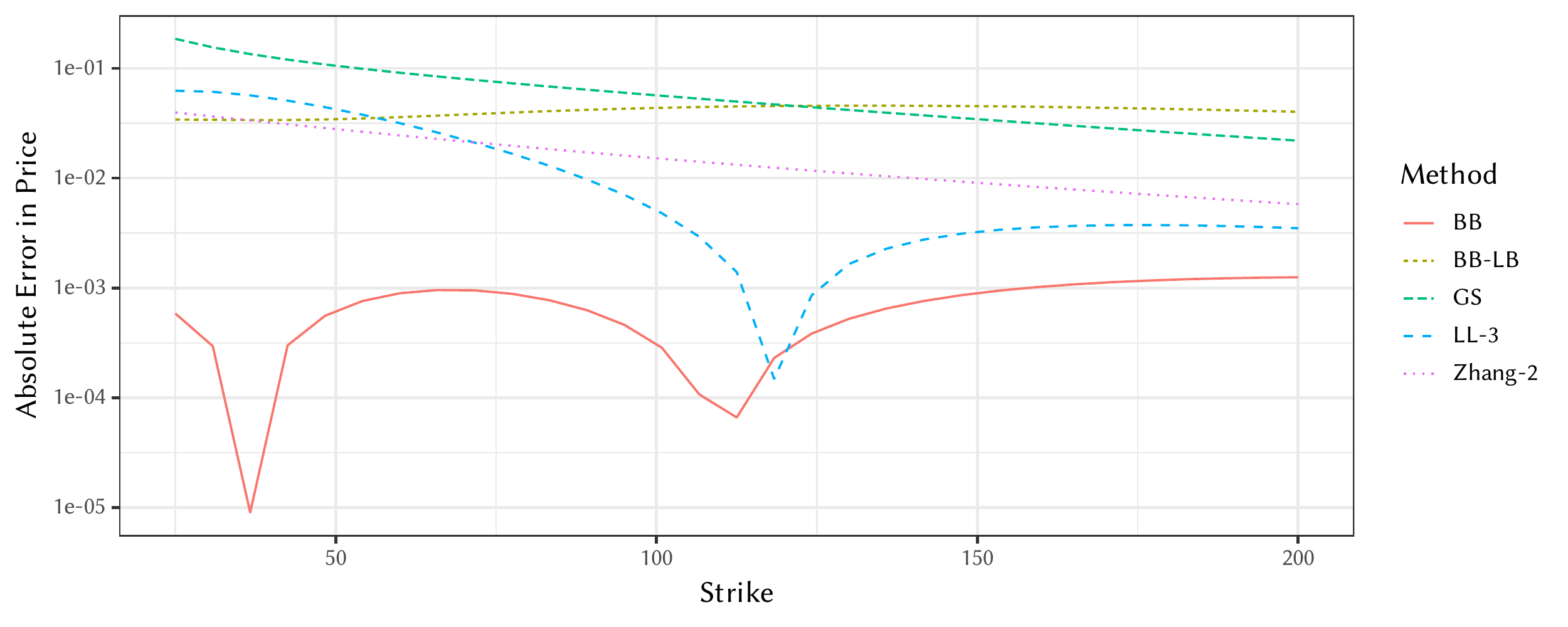}		
		\caption{Error of the various approximation in the case of two large dividends $\alpha_i=25$ and high volatility $\sigma=80\%$, for an option of maturity 1 year, varying the strike price.\label{fig:zhang_extreme}}
	\end{center}
\end{figure}

We now explore the extreme settings from \cite{zhang2011fast} where the dividend amount is very large, and the volatility is high. 
Figure \ref{fig:zhang_extreme} shows the absolute error in the price of a call option of maturity 1 year, on an asset of spot price $S=100$ paying two dividends of size $\alpha_i=25$ at $t_1=0.3$ and $t_2=0.7$. The interest rate is $r=5\%$ and the volatility is $\sigma=80\%$.
Again, the basket approach BB is the most accurate one, and is significantly more accurate than any other approximations. The lower bound approximation BB-LB is slightly more accurate than GS
\subsection{Performance}
The basket approximation from \citet{deelstra2010moment} requires a one-dimensional numerical integration and as such is slower than the alternatives. We use an adaptive Simpson quadrature for the integration, with an error tolerance set to $10^{-7}$, such that the result of the quadrature may be considered exact for all practical purposes.
For a large number of dividends, this basket approximation will however be faster than third-order approximations as the computational complexity is in $O(n^2)$ where $n$ is the number of dividends vs. $O(n^3)$ for third-order methods.

\begin{table}[h]
	\centering{
		\caption{Time to price 1000 vanilla options with 100 dividends yo the option maturity.\label{tbl:cash_perf}}
		\begin{tabular}{l c c c c c c}\toprule
 Approximation & BB-LB &GS & Zhang-2 & BB-Ju & LL-3 & BB\\\midrule
 Time (s)   & 0.10  & 0.12 & 0.15 & 2.08  & 4.74   & 10.20\\\bottomrule
\end{tabular}
}
\end{table}

      
While the Deelstra-based approximation is around 70 times slower than other second-order methods, it is unlikely to be a real bottleneck in a financial system. Instead of the one-dimensional integration, it is also possible to use the corresponding lower-bound approximation (named "BB-LB" in Table \ref{tbl:cash_perf}), where no numerical integration is required. The method is then the fastest, but its accuracy becomes similar to the least accurate second-order method, GS.

%

\section{Conclusion}
We have shown that any  basket option approximation for the standard Black-Scholes model may be directly used to price a European option on single aasset under the piecewise-lognormal model.

For a single cash dividend, the Curran based basket approximation leads to prices as accurate as advanced second-order or third-order cash dividend specific approximations. In the context of many dividends, the basket approach is found to be the most accurate one while the computational cost evolves as the square of the number of dividends, in similar fashion as the second-order cash dividend approximations. The need for a numerical integration makes it slower than other existing cash-dividend approximations in general. This is however unlikely to be a real issue in a financial system. The lower bound basket approximation, while less accurate, does not require a numerical integration and was found to be of similar accuracy as the approximation of Sahel and Gocsei, while being faster the the latter.

The technique presented in this paper may be easily extended to to price vanilla basket options under the piecewise lognormal model through the classic Black-Scholes basket option approximations, something that specific discrete dividend approximations typically do not allow.

\externalbibliography{yes}
\bibliography{cash_dividend_basket}
\end{document}